\RequirePackage{pdf14}
\PassOptionsToPackage{table}{xcolor}
\documentclass[sigconf,review]{acmart}
\usepackage[a-1b]{pdfx}
\usepackage{changepage} 

\AtBeginDocument{%
  \providecommand\BibTeX{{%
    \normalfont B\kern-0.5em{\scshape i\kern-0.25em b}\kern-0.8em\TeX}}}

\usepackage{booktabs}
\usepackage{csquotes}
\usepackage{graphicx}
\usepackage{verbatim}
\usepackage{fancybox}
\usepackage{multirow}
\usepackage{tabularx}
\usepackage{cleveref}

\begin{document}

\title{Industry Experiences with Large-Scale Refactoring}

\author{James Ivers}
\email{jivers@sei.cmu.edu}
\affiliation{
  \institution{CMU Software Engineering Institute}
  \city{Pittsburgh}
  \state{PA}
  \country{USA}
}

\author{Robert L. Nord}
\email{rn@sei.cmu.edu}
\affiliation{
  \institution{CMU Software Engineering Institute}
  \city{Pittsburgh}
  \state{PA}
  \country{USA}
}

\author{Ipek Ozkaya}
\email{ozkaya@sei.cmu.edu}
\affiliation{
  \institution{CMU Software Engineering Institute}
  \city{Pittsburgh}
  \state{PA}
  \country{USA}
}

\author{Chris Seifried}
\email{cgseifried@sei.cmu.edu}
\affiliation{
  \institution{CMU Software Engineering Institute}
  \city{Pittsburgh}
  \state{PA}
  \country{USA}
}

\author{Christopher S. Timperley}
\email{ctimperley@cmu.edu}
\affiliation{%
  \institution{Carnegie Mellon University}
  \city{Pittsburgh}
  \state{PA}
  \country{USA}
}

\author{Marouane Kessentini}
\email{kessentini@oakland.edu}
\affiliation{%
  \institution{Oakland University}
  \city{Rochester}
  \state{MI}
  \country{USA}
}

\renewcommand{\shortauthors}{Ivers, et al.}

\begin{abstract}
  Software refactoring plays an important role in software engineering. Developers often turn to refactoring when they want to restructure software to improve its quality without changing its external behavior. Studies show that small-scale (floss) refactoring is common in industry and can often be performed by a single developer in short sessions, even though developers do much of this work manually instead of using refactoring tools. However, some refactoring efforts are much larger in scale, requiring entire teams and months of effort, and the role of tools in these efforts is not as well studied. In this paper, we report on a survey we conducted with developers to understand large-scale refactoring, its prevalence, and how tools support it. Our results from 107 industry developers demonstrate that projects commonly go through multiple large-scale refactorings, each of which requires considerable effort. While there is often a desire to refactor, other business concerns such as developing new features often take higher priority. Our study finds that developers use several categories of tools to support large-scale refactoring and rely more heavily on general-purpose tools like IDEs than on tools designed specifically to support refactoring. Tool support varies across the different activities, with some particularly challenging activities seeing little use of tools in practice. Our study demonstrates a clear need for better large-scale refactoring tools and an opportunity for refactoring researchers to make a difference in industry. The results we summarize in this paper is one concrete step towards this goal.
\end{abstract}



\maketitle

\section{Introduction}
Refactoring is defined as restructuring software to improve its quality without altering its external behavior~\cite{opdyke1992refactoring}. The need to restructure software can come from such diverse goals as improving software quality, migrating to new platforms like cloud, containerizing software for DevOps, incorporating new technologies, or extracting capabilities for strategic reuse. Many of these scenarios involve refactoring at a large scale and imply broad changes to the system that cannot be accomplished through local code changes. This paper focuses on these larger refactoring efforts, which we refer to as \textbf{large-scale refactoring (LSR)} and is the first study of large refactoring efforts (mean estimated efforts greater than 1500 staff days) from multiple industry organizations. 

Such broad changes, however, are often hindered by software complexity and require labor intensive efforts to complete.  Consequently, developers continue to desire more time to conduct refactoring activities ~\cite{dagstuhlDig14}, often combine refactoring with new feature development to gain approval to proceed ~\cite{Kim2012FSE}, and seek more tool support while not trusting tools that are available as previous empirical studies reveal ~\cite{Murphy2008tools,kim2014tse}.

Well known refactoring types described by Martin Fowler (e.g., rename, move function, extract class) are frequently used by developers ~\cite{Fowler1999refactoringBook}. Integrated development environments (IDEs) like IntelliJ IDEA, Eclipse, VS Code, and Visual Studio all include features that change code to apply primitive refactoring types as directed by users.
However, these tools have varying levels of acceptance by developers. For example, in a study done with 328 Microsoft developers, Kim et al. found that 86\% refactor manually, with minimal use of features that implement the refactoring types they intend to use \cite{kim2014tse}. These results mirror results of earlier studies showing that in industry manual efforts dominate use of available tool support for refactoring ~\cite{Murphy2006eclipse, Vakilian2012refactorings, murphy-hill07, Murphy2008tools, murphy2012tse}. 

These well-established refactoring types are also used in large-scale refactoring \cite{kim2014tse,hyrum2013,hyrum2019}. Prior studies show use of such refactorings to address evolution of APIs~\cite{Dig2005api,Weissgerber2006api,Kim2011api}, design ~\cite{bavota2014rss}, and architecture ~\cite{Arcelli2018easier,bavota2014rss,lin2016interactive,mkaouer2016use,terra2012recommending,Zimmermann2017arch}. However, while large-scale refactoring is anecdotally performed in industry as demonstrated by these studies, it is not explicitly studied as part of software evolution or refactoring tool support. 

To understand how developers engage with large-scale refactoring and how they use tools to support different activities involved, we conducted a developer survey. Our findings confirm existing research on the challenges of smaller-scale refactoring activities. However, our results demonstrate that when it comes to tools that developers use to perform large-scale refactoring, developers use several categories of tools beyond those that implement refactorings in code. Tool support varies across the different activities that are involved in large-scale refactoring, with some particularly challenging activities seeing little use of tools in practice. While developers broadly agree that better tools are desired, they vary in the activities and degree of intelligence they want in tools.

Our contributions include the following:
\begin{itemize}
\item Our study is the first to specifically focus on large-scale refactoring, demonstrate its prevalence with industry empirical data, and position it as part of refactoring research and tool agendas. This contribution provides empirical data that challenges the assumptions that research should mostly focus on small-scale (floss) refactoring.  
\item We identify common reasons for deciding to perform or forgo large-scale refactoring. We further identify common consequences of forgoing such refactoring, which adds a missing perspective to industry's motivation in engaging in large-scale refactoring.
\item We identify which refactoring activities are most challenging, time consuming, and see the greatest and least use of tools. We further identify the categories of tools that are used today, which further improves our understanding of gaps in refactoring tool support.
\item Lastly, we share our data.
\end{itemize}

This paper is organized as follows. Section \ref{sec:background} summarizes research in different kinds of refactoring and the refactoring process and introduces large-scale refactoring. Section \ref{sec:methodology} describes our study approach. Section \ref{sec:results} presents our analysis results, whose implications are discussed in Section \ref{sec:discussion}. Section \ref{sec:conclusion} presents our conclusions.

\section{Background}
\label{sec:background}
Refactoring is a complex activity involving problem recognition, problem analysis, decision making, implementation, and evaluation ~\cite{Haendler2018process}. Despite increasing research interest in providing tool support in refactoring in terms of design, composition, and decision making~\cite{Mens2004survey}, developers continue to be reluctant to adopt automated refactoring support ~\cite{jetbrainsSurvey2021}. 

Murphy-Hill and Black ~\cite{Murphy2008tools} introduced two different notions of refactoring; the need to continually tweak code while making other changes (floss refactoring) and infrequent, but focused changes to improve unhealthy code (root-canal refactoring). Floss and root-canal refactoring are primarily differentiated by the nature of changes made -- floss refactoring intermingles refactoring with other changes, like feature development, while root-canal is almost entirely about refactoring.  Root-canal refactoring is also typically described as correcting unhealthy code, emphasizing quality improvements rather than other motivations.  Multiple empirical studies that use analysis of commit histories or IDE usage have found more evidence of floss refactoring than evidence of root-canal refactoring \cite{murphy2012tse,liu2012reftactics,sousa2020archref}. Murphy-Hill et al's study \cite{murphy2012tse} further suggests that "studies should focus on floss refactoring for the greatest generality."

Tools that implement refactorings in code are available for many popular programming languages through IDE context menu options that provide a list of available refactoring types from which developers choose, such as those included in IntelliJ IDEA, Eclipse, VS Code, and Visual Studio. According to a recent JetBrains survey, developers do in fact refactor their code every week or even almost every day and refactoring 
sessions often last an hour or longer. Despite this tool support, developers frequently refactor their code manually, often due to a lack of trust in what tools would do ~\cite{jetbrainsSurvey2021}. Furthermore, studies analyzing GitHub contributions reveal that refactorings are driven more often by changing requirements than by code smells ~\cite{Silva_2016}. There is clearly an aspect of refactoring that is broader in scope than the local code improvements that its original definition and floss refactoring recommendations implied.  

We define \textbf{large-scale refactoring} as restructuring software, without introducing functionality, for the purpose of improving non-functional quality or changing architecture. Large-scale involves either pervasive changes across a codebase or extensive changes to a substantial element of the system (e.g., greater than 10k LOC). Large-scale refactoring often involves a substantial commitment of resources, requiring management approval. One example is the need to partition legacy monoliths into smaller pieces to create separately deployable, scalable, and evolvable units. Another is restructuring interfaces and communication patterns to enable replacement of a legacy feature by an improved or less proprietary alternative. 

Large-scale refactoring is closely related to root-canal refactoring in that both focus on structural improvements that are not intermingled with other changes. However, we distinguish it from the common use of root-canal refactoring in scale and motivation. Many examples of root-canal refactorings in the literature do not represent particularly large efforts that require management support and significant commitment of resources and focus on system wide code quality improvements. Using this distinction in our survey, we were able to capture data for refactoring efforts that were estimated to require a mean of more than 1500 staff days of effort. Such large refactoring efforts are often motivated by broader business concerns than quality improvement and we wanted to be more inclusive of other business motivations. 

Architecture refactoring can be considered a form of large-scale refactoring. The work of Lin et al. is closest in its motivation to the large-scale refactoring notion that our survey investigated ~\cite{lin:fse2016}, but their work is focused on developing a research tool that relies on recommendations of a limited number of refactorings. Other work in architecture refactoring mostly addresses code and architecture smell detection, which often focus narrowly on quality symptoms to hint at opportunities for architecture level changes ~\cite{sousa2020archref, lucaArchSmellRefactIWR18}. Our study, in contrast, takes a broad perspective on the range of activities and supporting tools from the perspective of developers performing large-scale refactoring in industry. 

A recent study of how software developers make decisions proposed a decision-making framework for refactoring ~\cite{Leppanen2015framework}. They found stages of decision-making that consist of a pain zone that triggers the decision of refactoring, situation analysis, refactoring planning, refactoring implementation, and follow up to assess the effort. Factors that lead to decision making are influenced by scale. More recently, Haendler and Frysak~\cite{Haendler2018process} provided a theoretical perspective on applying concepts from decision-making research to deconstruct the refactoring process. They provide a more general interpretation of the software maintenance process~\cite{Kitchenham1999maintenance} and different refactoring stages~\cite{Leppanen2015framework}: problem recognition, problem analysis, decision-making, implementation, evaluation. Furthermore, the model introduces a second dimension to account for the primary decisions in refactoring at management and operational levels: whether to refactor, what to refactor, how to refactor. The authors then group the many tools and techniques available for refactoring by the following characteristics: smell detection and refactoring recommendation tools, code-quality and design-critique tools, refactoring tools, technical debt management and analysis tools, automated regression testing frameworks, and documented knowledge on refactoring rules. Our survey also reveals tools used across these categories and confirms that the support is not ideal.

These studies commonly point out that the refactoring process consists of activities that span several decision-making stages as well as activities along the software development lifecycle.  Abid et al. recently completed a literature survey spanning 30 years of refactoring research that emphasized a lifecycle view of refactoring ~\cite{abid202030}. In our survey, we build on these studies and focus on the following activities.
\begin{itemize}
\item Determining where changes were needed
\item Choosing what changes to make 
\item Implementing the changes
\item Generating new tests
\item Migrating existing tests
\item Validating refactored code (inspection, executing tests, etc.) 
\item Re-certifying refactored code (common to industry in regulated domains)
\item Updating documentation
\end{itemize}

Through the rest of the paper we make use of these activities to understand the prevalence of large-scale refactoring, its challenges, and gaps in existing tools that support large-scale refactoring activities. Our survey is not the first survey study to focus on refactoring. Kim et al. ~\cite{kim2014tse} conducted a survey study with 328 Microsoft developers in 2014 to understand the benefits of refactoring and developer perceptions. Their conclusions included that the definition of refactoring in practice is broader than behavior-preserving program transformations and include system wide changes. In addition, they showed that developers need various types of refactoring support beyond the refactoring features provided by IDEs. More recently, Golubev et al. surveyed 1183 IntelliJ users and reaffirmed that many developers do not trust automated refactoring features. Our survey study is similar in its methodology to Kim and Golubev; however, it differs in its motivation and is the first to explicitly target large-scale refactoring, establish it explicitly as a distinct refactoring category, and provide insights into the tooling challenges it entails. 

\section{Methodology}%
\label{sec:methodology}

Our goals in this study include assessing how developers perform large-scale refactoring and understanding the tools they use to support the process and their shortcomings. To achieve these goals, we ask the following research questions. 

\textbf{RQ1:} Is large-scale refactoring common in industry and what drives decision making?

\textbf{RQ2:} How do developers use tools to aid their large-scale refactoring efforts?

\textbf{RQ3:} What tools and support, if any, do developers desire to aid their large-scale refactoring efforts?

In our first question we look for overarching business and technical goals, reasons why and why not to refactor, and risks and challenges associated with large-scale refactoring. Our other questions then focus on refactoring process activities, examine the role of tools to support these activities, and what kind of tools would better support these activities.

\begin{figure}
\begin{center}
\newcommand*\rot{\rotatebox{90}}
{\small  
\begin{tabular}{|l|l|}
\hline
\multirow{5}{*}{\rot{\textbf{RQ1}}}
 & $\bullet$ What were the business goals of the refactoring? \\
 & $\bullet$ Have you ever wanted to perform a large-scale refactoring \\
 & but were unable to? \\
 & $\bullet$ What consequences, if any, did you observe from not \\
 & performing the refactoring? \\
\hline
\multirow{3}{*}{\rot{\textbf{RQ2}}}
 & $\bullet$ What tools, if any, did you use to assist your large-scale \\
 & refactoring efforts? \\
 & $\bullet$ To what extent do you use tools for the following activities? \\
 \hline
\multirow{4}{*}{\rot{\textbf{RQ3}}}
 & $\bullet$ What kind of automation, if available, would have most \\
 & improved your large-scale refactoring? \\
 & $\bullet$ What are the strengths and weakness of the tools you used \\ 
 & to support large-scale refactoring? \\
\hline
\end{tabular}
}
\end{center}
\caption{A sample of our survey questions and their corresponding research question (RQ). 
}
\label{fig:survey}
\end{figure}%

\textbf{Survey Design}.
To answer our research questions, we performed an online survey of members of the
software engineering community between November, 2020 and February, 2021.
To ensure that we collected meaningful and informative results,
we followed several survey design best practices by explicitly deriving survey questions from our research questions, conducting a series of iterative pilot surveys on a representative population of sample respondents, and refining our survey design until reaching
saturation~\cite{principles-survey-research,survey-design-experiences}.
A sample of our survey questions is given in \Cref{fig:survey}. We used a branching design to elicit separate experiences in which participants had performed large-scale refactoring and those in which they had been unable to do so.
Those who had performed large-scale refactoring were presented questions related to the challenges, outcomes, and the extent to which tools supported the process. Those who were unable to do so answered questions as to why not and the consequences of not refactoring.  

\textbf{Recruitment}.
Our survey targeted an industry audience. We distributed our survey to members of the software engineering community via email (dlist: 7,700), LinkedIn (subscribers: 16,012), Twitter (followers: 5,383), research colleagues (for redistribution to their industry collaborators), and company internal technical interest groups. A total of 107 participants took part in the survey. 96\% of participants were software engineers and/or software architects (both of which we refer to as developers) and 74\% worked in industry. 79\% of the participants had 10+ years of experience (\autoref{fig:demographics}). These demographics demonstrate that our participants represent a wealth of collective industry experience, which helps to increase the confidence in our findings.

\begin{figure}
\begin{tabular}{lrr}
\toprule
\textbf{Years of experience} & \textbf{\#} & \textbf{\%} \\
\midrule
Less than three years & 6 & 6\% \\
Between three and ten years & 16 & 15\% \\
Ten or more years & 85 & 79\% \\
\bottomrule
\end{tabular}
\caption{Demographics of our 107 survey participants in terms of their years of experience in the software industry.}
\label{fig:demographics}
\end{figure}

\textbf{Qualitative Analysis}.
\label{sec:methodology:analysis}
To analyze responses to the qualitative parts of our survey, we used a descriptive coding approach~\cite{saldana2015coding}.
We first tagged each response to open-ended survey questions with one or more labels, known as codes, describing the topics of that response.
We then performed adjudication and code mapping to collapse our codes into a consistent set of categories.
Finally, we used axial coding to identify relationships between categories, and to identify a small number of overarching themes.
Throughout this process, we performed continual analysis, comparison, and discussion of data until reaching thematic saturation (i.e., no new perspectives, dimensions, or relationships were identified).

These responses reflect developers' perceptions and experiences of large-scale refactoring.
We report frequency of our coding of this data only to demonstrate the prevalence of themes in our data, not to suggest generalized conclusions. To allow others to understand the logic behind our analysis process, we also provide sample quotes throughout the paper.

\textbf{Study Artifacts}.
To promote further research and allow others to inspect and replicate our methodology and findings,
we provide a detailed audit trail of our study artifacts, which include our
survey questionnaire, recruitment materials, codebook, anonymized survey data, and the Jupyter notebook used to produce the figures in the paper.
Our study artifacts are available at: \hfill \break \url{https://github.com/ArchitecturePractices/lsr\_survey\_artifacts}.

\section{Results}%
\label{sec:results}

In this section, we report our analysis and key insights on the data from 107 responses. 

\subsection{RQ1: Is large-scale refactoring common in industry and what drives decision making?}
\label{sec:results:rq1}

Common perception is that business priorities and natural system evolution drive the need to conduct large-scale changes in industrial software. However, do we know whether developers resonate with the concept of large-scale refactoring and how frequently, if at all, they engage in conscious large-scale refactoring activities? We wanted to understand the business and technical triggers, as well as the challenges surrounding the decision to perform or forgo large-scale refactoring and their consequences.

\begin{figure}[ht!]
    \centering
        \centering
        \includegraphics[width=1\columnwidth]{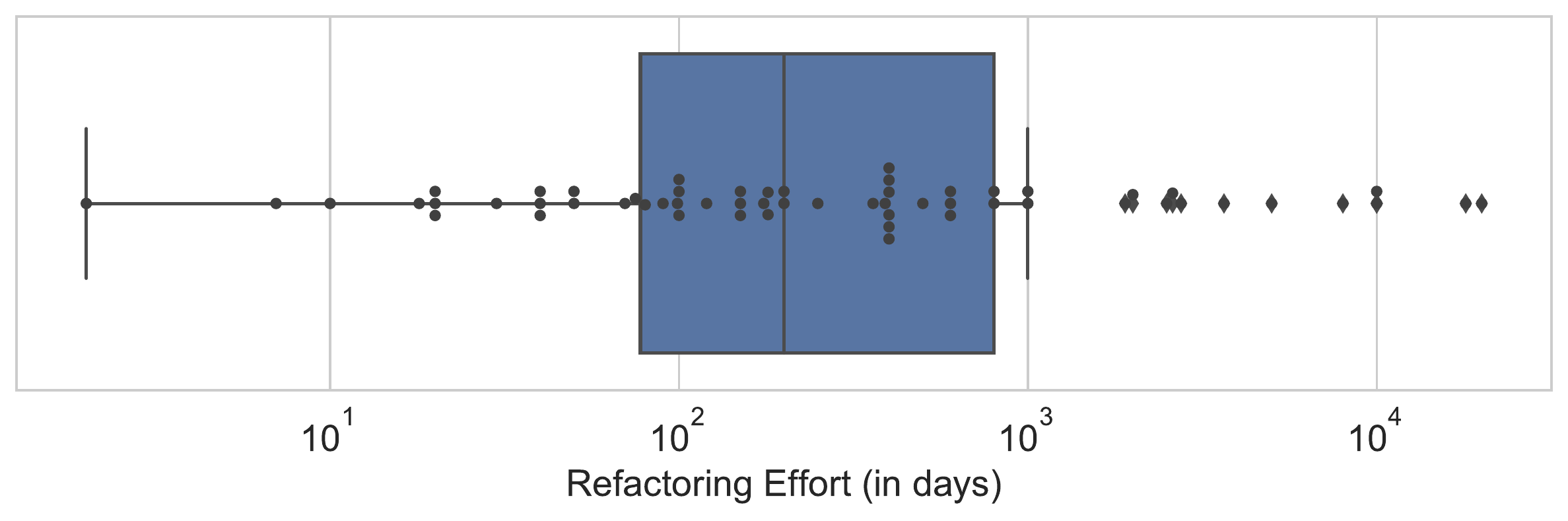}
        \caption{Estimated effort (in staff days) that teams required to complete their large-scale refactorings.}
        \label{fig:lsr_time}
\end{figure}

\textbf{Prevalence of LSR}.
82\% of respondents had participated in large-scale refactoring at least once. Of the 61\% who reported participating in large-scale refactoring more than once, 12\% of respondents had engaged in such refactoring five or more times. These refactorings were performed on significantly large systems (34\% were larger than 1M LOC and 38\% ranged from 100K-1M LOC) and consumed significant resources, ranging from 2 days to 20,000 staff days as shown in \autoref{fig:lsr_time}. 
Furthermore, 56\% of systems on which respondents had performed large-scale refactoring had undergone large-scale refactoring multiple times (16\% twice, 36\% three to five times, and 5\% more than five times).
Half of respondents reported that they are still working on the same system on which they had performed large-scale refactoring, 42\% of whom have worked on this system for more than five years. The release frequencies for these systems ranged from several times a month (25\%) to several time a year (49\%).
These results confirm the common wisdom that industrial software systems go through major changes and support the conclusion that organizations do commonly conduct large-scale refactoring. 

\begin{figure}[ht!]
    \centering
        \centering
        \includegraphics[width=1\linewidth]{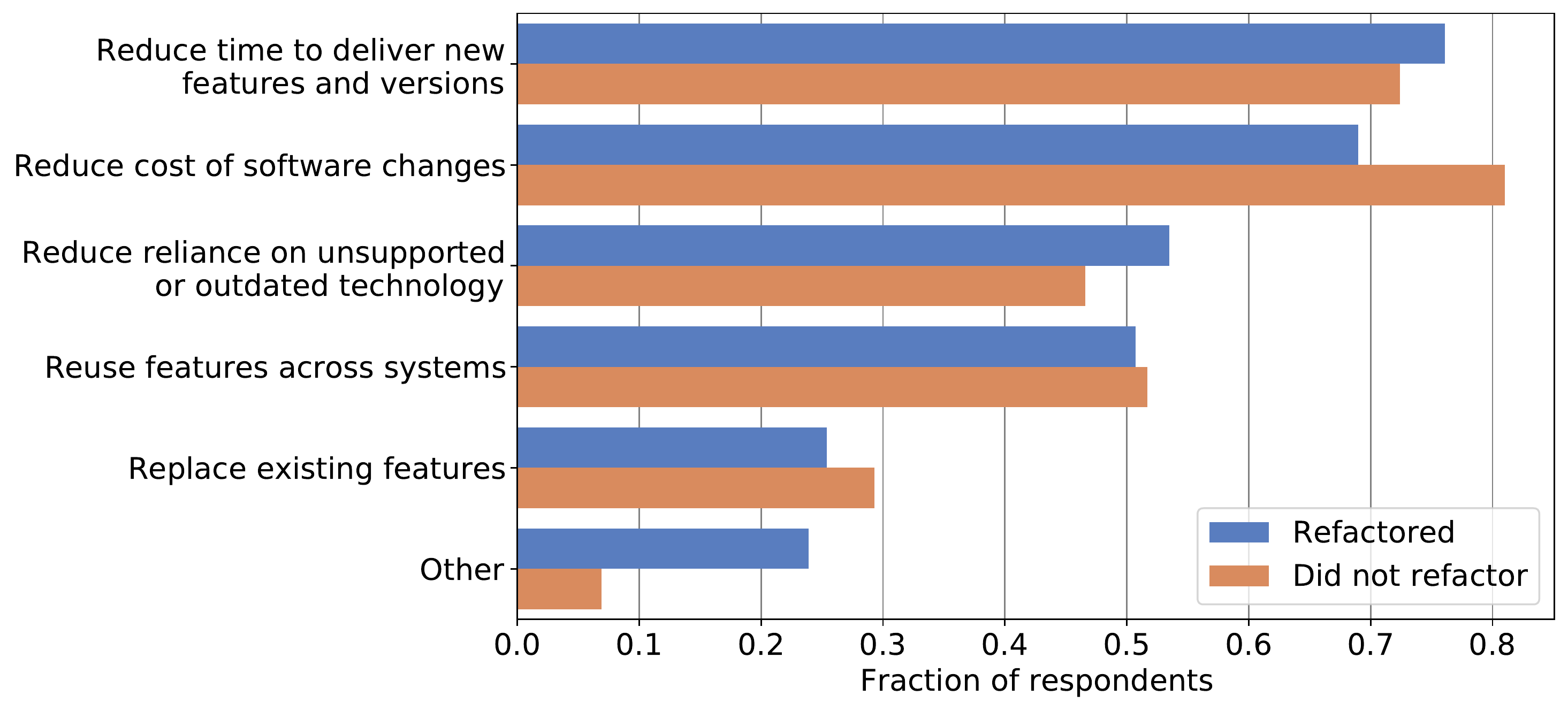}
        \caption{Business reasons for large-scale refactoring.}
        \label{fig:bizreasons}
\end{figure}

\begin{figure}[ht!]
        \centering
        \includegraphics[width=1\linewidth]{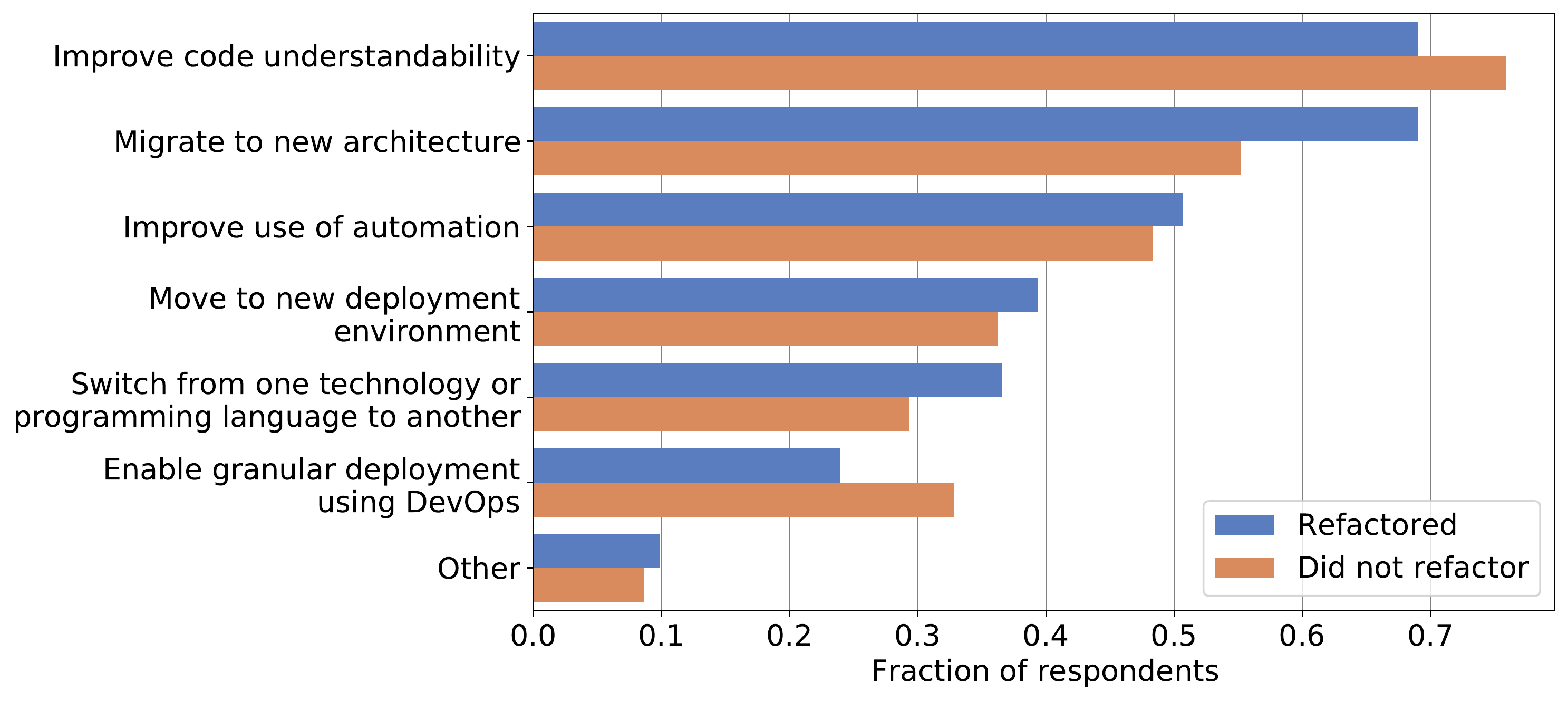}
        \caption{Technical reasons for large-scale refactoring.}
        \label{fig:techreasons}
\end{figure}

\textbf{Reasons for LSR}.
Reducing cost of change and time to deliver were expressed as top business reasons to refactor by our respondents who both had the opportunity to refactor and wanted to refactor but could not (\autoref{fig:bizreasons}). As for technical reasons for refactoring, improving understandability and migrating to a new architecture had similarly top occurrences (\autoref{fig:techreasons}).

Our analysis revealed an interesting relationship between the top business and technical reasons: 78\% of those who reported that reducing cost of change was a business reason to refactor also reported improving code understandability as a technical reason to refactor. Among those who did undertake large-scale refactoring, 70\% reported both improving code understandability and migrating to a new architecture as top technical reasons to refactor. These results further demonstrate the relationship between the kind of architectural change that requires large-scale refactoring and the potential impact of such change on business goals. 

\begin{figure}[ht!]
        \centering
        \includegraphics[width=1\linewidth]{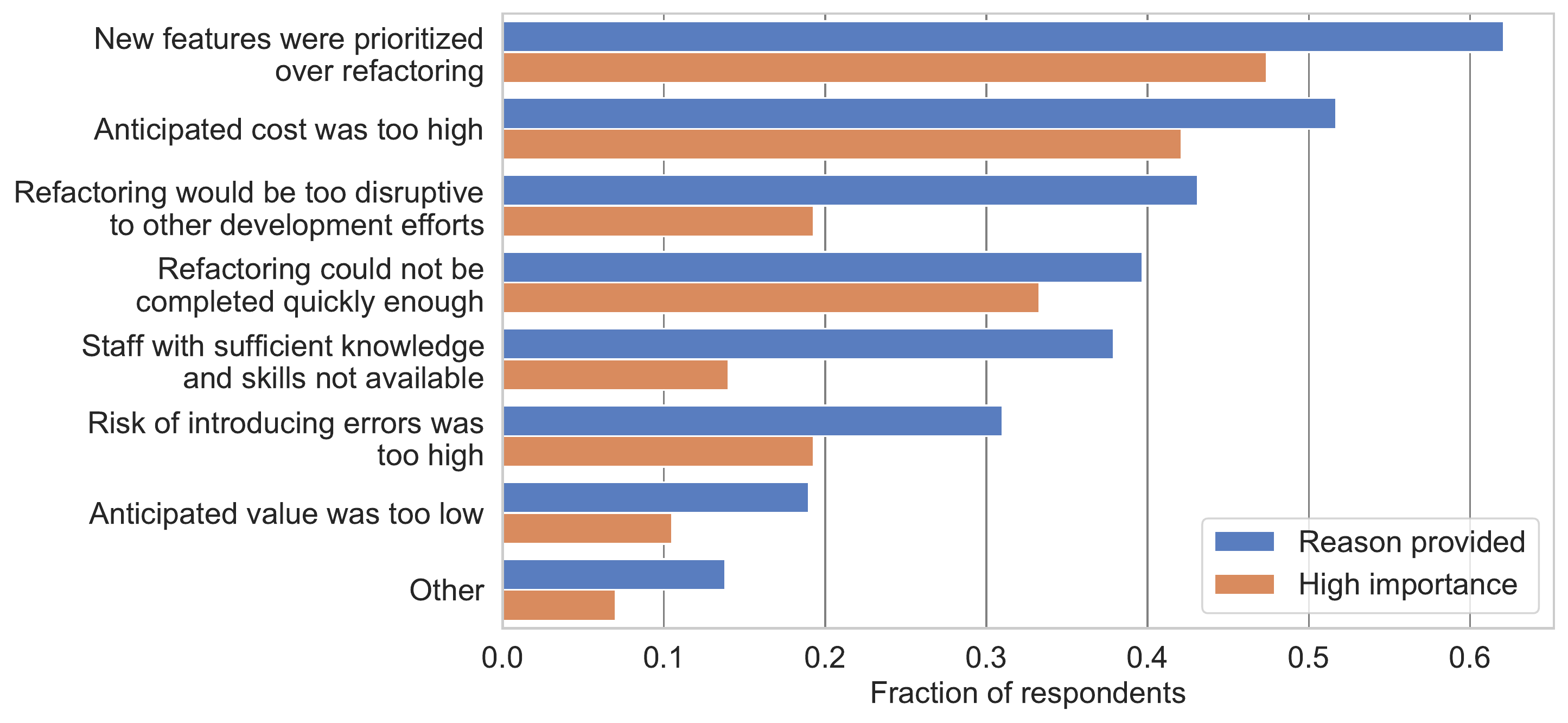}
        \caption{Reasons why organizations forgo large-scale refactoring.}
        \label{fig:reasonswhynot}
\end{figure}

\textbf{Forgoing LSR}.
Having established that industry systems undergo multiple large-scale refactorings, we looked at how often organizations had wanted to perform refactoring but had decided not to do so. 71\% of respondents reported that there were occasions that they wanted to conduct large-scale refactoring, but were unable to do so. Sharma's study reported a similarly high portion of respondents (76\%) identifying prioritization of features over refactoring as an obstacle to undertaking refactoring \cite{Sharma2015industry}.
The reasons for deciding not to perform large-scale refactoring centered around opportunity cost (new features were prioritized and anticipated cost was too high) as the most important reasons. 35\% of respondents reported both as driving reasons, indicating that when resources are scarce, new features are commonly preferred over other investments. Interestingly, among the reasons to not refactor, only 6\% of the participants indicated that the anticipated value of refactoring was too low (\autoref{fig:reasonswhynot}). 

\textbf{Consequences of Forgoing LSR}.
Given business realities, these results are not surprising and they align with previous refactoring research ~\cite{kim2014tse,Sharma2015industry}. When resource constraints (especially time and cost) force choices, new features are prioritized over refactoring. However, there are consequences to not performing needed refactoring, as our participants reported through open ended questions.  When we analyzed these responses through a coding exercise (\autoref{table:consequences}), we found that the most common long term consequences were related to inability or slowing paces of delivering new features (56\%). Instances of deteriorating internal (54\%) and external (32\%) quality were often accompanied by references to increasing operating or development costs, which are expected consequences of quality deterioration. 90\% of respondents reported delivery and/or internal quality problems, both of which reflect slowing development velocity, as consequences of not refactoring.  These consequences undermine the perceived opportunity to divert resources from refactoring to new features.

The consequences that participants shared also clearly exemplify the need for large-scale refactoring.
\begin{itemize}
\item\textit{We are stuck on outdated technologies. It is difficult to keep up with the "startup" companies that provide features that we are not able to create on the old tech stack.}
\item \textit{...modernization cycle was held back by 4 years....maintenance cost stayed high....cost to implement, deploy, and validate continue to increase}. 
\item \textit{Feature delivery took longer as it required changes to multiple parts of the system.} 
\end{itemize}
Not surprisingly, long term consequences of not refactoring include jeopardizing the top priority business concern of reducing time to deliver new features, as well as increased cycle time and costs. 

\begin{table}
\begin{tabularx}{\columnwidth}{lXr}
\toprule
\textbf{Category} & \textbf{Description} & \textbf{\%} \\
\midrule
Delivery & Slow feature delivery, inability to \newline develop features & 56\% \\
Internal quality & Low productivity, duplicated code, non-bug design flaws & 54\% \\
External quality & Degraded user experience, bugs, \newline performance issues & 32\% \\
Staffing & Low morale, increased onboarding time, difficulty hiring or retaining staff & 22\% \\
\bottomrule
\end{tabularx}
\caption{Consequences of forgoing large-scale refactoring, by fraction of respondents reporting each category.}
\label{table:consequences}
\end{table}
\medskip
\textbf{Findings}
\begin{itemize}
\item 82\% of respondents had performed large-scale refactoring.  Of the systems on which they had performed large-scale refactoring, 57\% had undergone multiple large-scale refactorings.
\item Large-scale refactorings are substantial efforts. 71\% had refactored systems of at least 100K LOC. The mean time to complete refactoring was estimated at more than 1500 staff days.
\item Forgoing large-scale refactoring is also common in industry, as 71\% of respondents had wanted to perform refactoring but were unable to do so.  
\item While prioritizing new features over refactoring was the most common reason for forgoing large-scale refactoring, 56\% of respondents reported the inability or slowing pace of delivering features as a consequence of forgoing refactoring.
\end{itemize}

\subsection{RQ2: How do developers use tools to aid their large-scale refactoring efforts?}%
\label{sec:results:rq2}

Refactoring has been a familiar concept to developers for decades ~\cite{Fowler1999refactoringBook}, but adoption of tools to support refactoring remains less common \cite{murphy2012tse,kim2014tse}. While studies have focused more on support for low-level refactoring than on large-scale refactoring, a study by Kim et al. included an analysis of interviews with a team that had performed system-wide refactoring on a very large system \cite{kim2014tse}. Their analysis indicates that refactoring at this scale involves far more than applying low-level refactorings. Instead, refactoring involved understanding the system, performing dependency analysis, creating a desired architecture structure, performing multiple gate checks, educating other developers, and developing custom refactoring tools. We sought to understand whether the kinds of tools used in large-scale refactoring differ from those used in other refactoring, the different activities involved in refactoring, and how those tools support those activities.

\begin{figure}[ht!]
        \centering
        \includegraphics[width=1
        \linewidth]{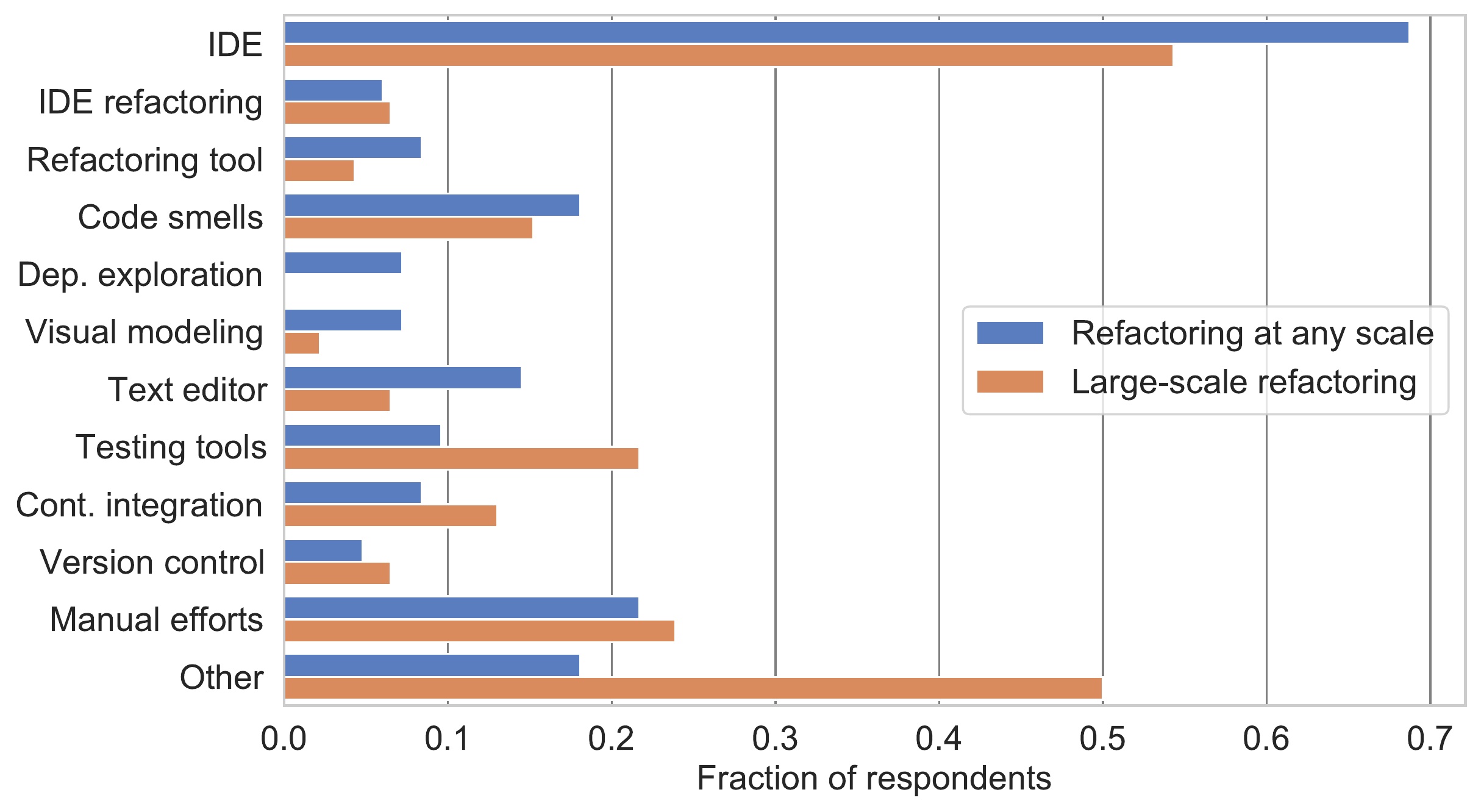}
        \caption{Categories of tools used to support refactoring.}
        \label{fig:tools-used}
\end{figure}

\textbf{Tools Used}.
We used two open ended questions to collect a list of tools that respondents used for refactoring at any scale and for large-scale refactoring. We used coding to categorize each tool into one of the categories shown in \autoref{fig:tools-used}, which contrasts the fraction of respondents using at least one tool in each category for refactoring at any scale with that for large-scale refactoring. There is little difference between the fraction of respondents using tools for large-scale refactoring and refactoring at any scale for most tool categories.  The exceptions are IDEs and text editors (greater use in refactoring at any scale), testing tools (greater use in large-scale), and other tools (much greater use in large-scale). The other tools category includes custom scripts and tools on which custom tools were likely built (static code analyzers and abstract syntax trees).

The most commonly used category of tool is the IDE; more than half of all respondents reported using IDEs for refactoring (68.7\% for any scale and 54.3\% for large-scale). In contrast, fewer than 10\% of respondents reported using tools that are designed specifically for refactoring like ReSharper and JDeodorant (8.4\% for any scale and 4.3\% for large-scale) or called out refactoring features of IDEs (6\% for any scale and 6.5\% for large-scale). The portion of tools falling into the other category was substantially higher for large-scale refactoring (50\%) than for refactoring at any scale (18\%).   

\begin{figure}[ht!]
        \centering
        \includegraphics[width=1\linewidth]{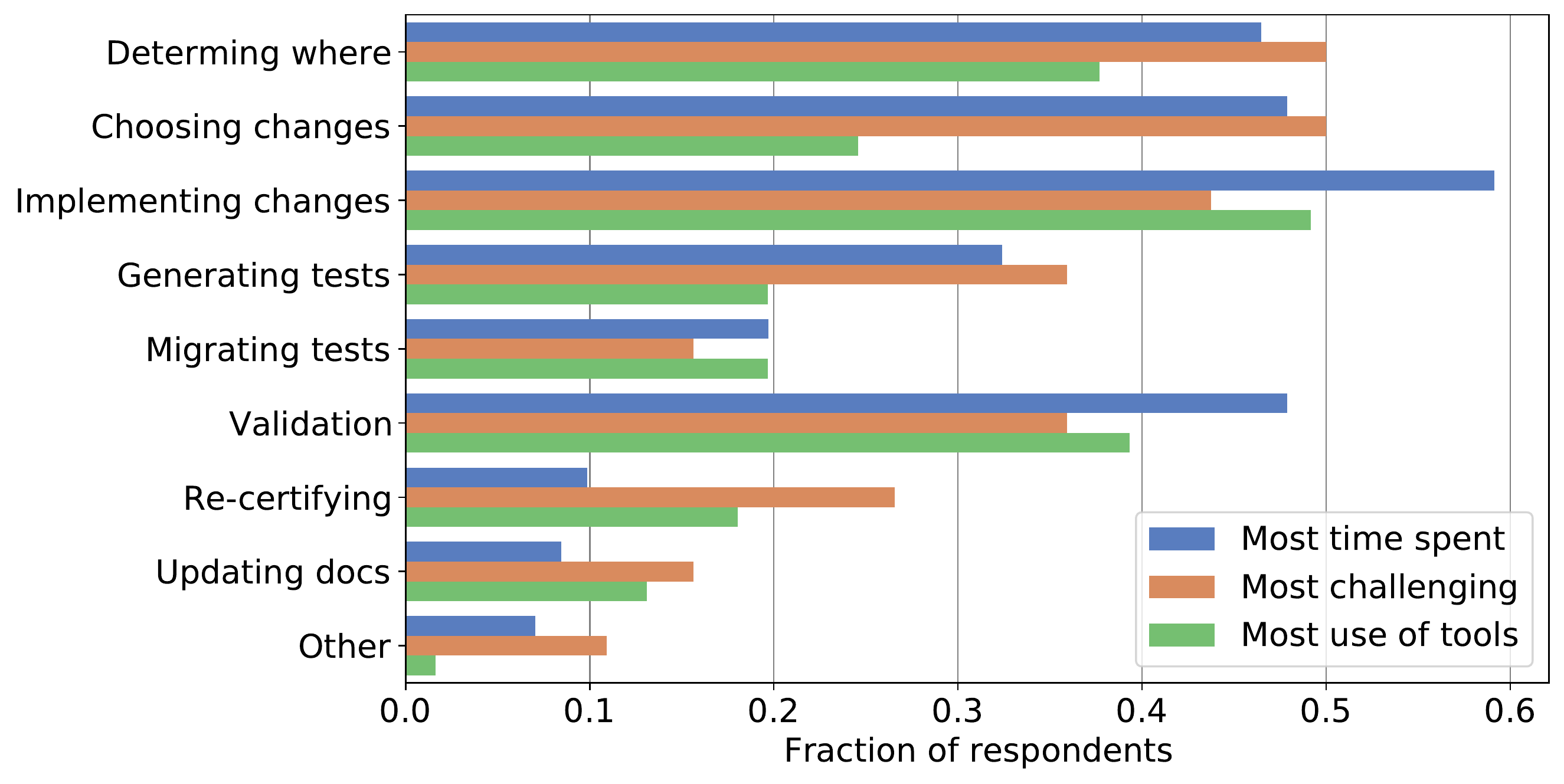}
        \caption{Refactoring activities that take the most time, are most challenging, and make the most use of tools.}
        \label{fig:activities_most}
\end{figure}

\textbf{Refactoring Activities}.
We next looked at the work that respondents perform as part of large-scale refactoring activities. We listed the refactoring activities found in \autoref{fig:activities_most} and asked respondents to report how much time they spent in each, how challenging they found each, and the extent to which they used tools for each. \autoref{fig:activities_most} shows the fraction of respondents reporting each activity in the positive for each question (i.e., most time spent, most challenging, and extensive use of tools). The top three activities in terms of what taking the most time, being the most challenging, and making the greatest use of tools all come from these four activities: (1) determining where changes are needed, (2) choosing what changes to make, (3) implementing changes, and (4) validating refactored code.  

Respondents commonly reported choosing what change to make as most time consuming (48\%) and most challenging (50\%), but only 25\% reported it as making extensive use of tools. In fact, looking at the negative responses, 58\% report this activity as making the least use of tools. 
The activity for which respondents reported least use of tools was updating documentation (75\%), which was also commonly noted as taking the least time (63\%) and being least challenging (61\%).

\textbf{Tool Effectiveness}.
Ideally, there is a relation among how much tools are used for an activity, how much time it takes, and how challenging it is.  Highly effective tools can dramatically reduce the time spent and the perceived challenge. Activities that remain highly challenging and time consuming can suggest shortcomings in or under-use of tools.
We compared the tools that respondents used for large-scale refactoring (\autoref{fig:tools-used}) with the time spent, challenge, and extent of tool use for refactoring activities (\autoref{fig:activities_most}). We also applied our judgment regarding the degree of support each category of tool provides (based on the specific tools listed by respondents) to each activity as additional context. 

Respondents report that determining where changes are needed and implementing changes as highly challenging activities (50\% and 43.8\%) for which they extensively use tools (37.7\% and 49.2\%), and yet both take significant time (46\% and 59\%). Of note, respondents report relatively little use of tools that are specifically designed to support these tasks. Refactoring, code smell analysis, and dependency exploration tools better support determining where changes are needed, but are used by only 4.3\%, 15.2\%, and 0\% of respondents. Refactoring tools and IDE refactoring features better support implementing changes, but are used by only 4.3\% and 6.5\% of respondents. Both indicate that while respondents use tools extensively for two of the three most challenging activities, they rely more heavily on general purpose tools like IDEs and manual effort than on tools specifically designed for refactoring.

\medskip
\textbf{Findings}
\begin{itemize}
\item Significantly more respondents use general-purpose tools like IDEs for large-scale refactoring (54.3\%) than use tools designed specifically for refactoring (less than 10\%). 
\item 50\% of respondents performing large-scale refactoring report use of other tools, which are dominated by custom tools, scripts, and packages on which they build their own tools.
\item Choosing which changes to make is one of the most challenging and time consuming activities, while also one of the activities for which developers make the least use of tools.
\item Updating documentation sees the least use of tools, but is also the least challenging and time consuming activity.
\end{itemize}

\subsection{RQ3: What tools and support, if any, do developers desire to aid their large-scale refactoring efforts?}
\label{sec:results:rq3}
To understand what kinds of tools would aid large-scale refactoring, we looked at the challenges respondents faced during their refactoring, the strengths and weaknesses of current tools, and how respondents directly answered the question.

\textbf{Activity Challenges}.
After asking respondents to rate how challenging each refactoring activity was, we asked them through an open response question what made their most challenging activities challenging. \autoref{table:challenges} shows our coding of these responses. Unsurprisingly, the most common challenge is the poor quality of the software being refactored, a challenge that refactoring exercises inherit as a starting point.

The second most common challenge is the difficulty in understanding code and the implications of a change. One respondent emphasized this as \textit{The hardest part was gaining a conceptual grasp of the overall code structure, and code flow, and understanding how one basic change – no matter how simple it appeared on the surface – might create consequences throughout the system.} While this challenge is more dependent on the tools and processes used for refactoring, the starting quality of code can exacerbate it. Half of respondents reporting code comprehension as a challenge also reported poor code quality or lack of documentation as a challenge. A need for code comprehension often stems from inheriting code written by someone else. Most respondents reported that the software on which they had performed large-scale refactoring was relatively old when they started working on it (for 27\% it was already 5-10 years old and for 25\% it was already more than 10 years old).

Respondents reported challenges that closely relate to code artifacts more than twice as often (code quality and comprehension at 34\% and 26\%) as challenges that relate to making decisions (scoping refactoring and decision criteria at 15\% and 6\%).

\begin{table}
{\small  
\begin{tabularx}{\columnwidth}{lXr}
\toprule
\textbf{Category} & \textbf{Description} & \textbf{\%} \\
\midrule
Code Quality & Poor quality of code being refactored, excessive dependencies that complicate changes & 34\% \\
Comprehension & Difficulties in understanding code structure, flow, and possible side-effects & 26\% \\ 
Tests & Lack of tests to ensure behavior & 19\% \\
Communication & Need to persuade management and teammates, gaining user trust & 19\% \\
Scoping & Managing expectations, deciding how much refactoring to do & 15\% \\
Documentation & Poor documentation, unclear intent & 13\% \\
Techniques & Lack of well-defined refactoring techniques & 11\% \\
Decision Criteria & Choosing the right changes & 6\% \\
\bottomrule
\end{tabularx}
}
\caption{What made large-scale refactoring challenging, by fraction of respondents reporting each category.}
\label{table:challenges}
\end{table}

\textbf{Current Tools}.
We next looked participant responses to a question on the strengths and weaknesses of the refactoring tools that they currently use. \autoref{table:str_weak} shows our coding of these responses.  Less than half of respondents provided any strengths, while only three respondents provided only strengths. The top categories for reported strengths were modification (16\%, automation of changes) and planning what (12\%, identifying opportunities for refactoring). The top categories for reported  weaknesses were usability (33\%, learning curve, poor interfaces for tasks) and modification (21\%, lack of control over or unacceptable results from automated refactoring). This corroborates a finding of Pinto and Kamei's study, which identified usability as a key barrier to adoption of refactoring tools \cite{Pinto2013stackoverflow}.

Several responses directly contrasted available refactoring support for small-scale changes with needs for large-scale refactorings.  Examples include:
\begin{itemize}
    \item \textit{They address refactoring efforts at a component level. They don't address end to end scenarios and analysing dynamics. Todays tools I have used provide quite a lot indicators for increasing complexity and structure loss, but these are not enough to make large scale decision with reducing these effects leading to system failure.}
    \item \textit{The tools I use don’t offer any guides or hints related to large-scale refactoring. Their analysis features usually present only low level code smells that often don’t offer a considerable improvement in the quality of the software.}
    \item \textit{The tools I've got are too focused on munging text, or on refactoring that is syntactically simple enough that I don't really need help with it (maybe it saves time on typing, but typing time isn't the problem).}
\end{itemize}

\begin{table}
\begin{tabularx}{\columnwidth}{Xrr}
\toprule
\textbf{Category} & \textbf{Strengths} & \textbf{Weaknesses} \\
\midrule
Usability & 7\% & 33\% \\
Modification & 16\% & 21\% \\
Planning what to refactor & 12\% & 19\% \\
Analysis & 9\% & 16\% \\
Large-scale refactoring & 5\% & 16\% \\
Comprehension & 2\% & 16\% \\
Testing & 2\% & 5\% \\
Planning how to refactor & 0\% & 5\% \\
Scoping refactoring & 0\% & 5\% \\
\bottomrule
\end{tabularx}
\caption{Strengths and weaknesses of tools respondents use for refactoring, by fraction of respondents reporting each category.}
\label{table:str_weak}
\end{table}

\textbf{Desired Tools}.
Despite 80\% of respondents who had performed large-scale refactoring reporting having achieved their goals, the activity challenges (\autoref{table:challenges}) and the weaknesses of tools used (\autoref{table:str_weak}) point to room for improvement.
We asked participants what kind of tools would have most improved their experience. \autoref{table:wants} shows our coding of these open ended responses. The three most common categories focused directly on the code being refactored. Testing (46\%) focused on testing automation, modification (26\%) focused on automating code changes, and analysis (23\%) focused on understanding the code (e.g., static and data flow analyses). In contrast, tools that included recommending actions were much less common: planning what to refactor (9\%, recommending where changes are needed) and planning how to refactor (6\%, recommending specific changes). Pinto and Kamei's analysis of Stack Overflow questions on refactoring identified generating refactoring recommendations as a desirable feature at a similarly low number (13\%) \cite{Pinto2013stackoverflow}.  Fewer respondents expressed interest in tools that make decisions for them than in tools that act as directed by a developer, like performing requested analyses, making specified changes, and confirming the results of changes. 

This preference aligns with \autoref{table:challenges}, which summarizes what made refactoring challenging.  Challenges with code comprehension and tests align with top wants.  Decision criteria was the least common challenge, aligning with the lack of wants for tools that recommend changes.
However, this preference is somewhat at odds with where respondents report spending the most time in \autoref{fig:activities_most}.  Two of the four activities on which they spend the most time (implementing changes and validation) align with two of the top three wants (testing and modification).  The other two activities on which they spend the most time (determining where changes are needed and choosing the changes to make) align with two of the least common wants (planning what and how to refactor).  This may reflect a lack of trust in tools' ability to make good recommendations, as evidenced by the following comment:
\begin{itemize}
    \item \textit{I'm still highly skeptical that a tool that can effectively automatically suggest a collection of refactoring that would solve a specific problem can be written. Refactoring is highly contextual... 
    Until you can create a program that can figure out the context of the problem just by analyzing the structure of the code (i.e. make the tool read people's minds), I doubt such a tool will ever exist.}
\end{itemize}

\begin{table}
\begin{tabularx}{\columnwidth}{Xr}
\toprule
\textbf{Category} & \textbf{\%} \\
\midrule
Testing & 46\% \\
Analysis & 23\% \\
Modification & 26\% \\
Comprehension & 17\% \\
Planning what to refactor & 9\% \\
Build Automation & 9\% \\
Planning how to refactor & 6\% \\
\bottomrule
\end{tabularx}
\caption{What kinds of tools would improve large-scale refactoring, by fraction of respondents reporting each category.}
\label{table:wants}
\end{table}

When asked how useful respondents would find a tool that automatically suggests a collection of refactorings that would solve a problem that you specified, 73\% of respondents replied affirmatively. Regardless of any skepticism in what tools can do, responses to the question of what tools would help reflected a genuine, if sometimes plaintive need for help (e.g., \textit{Any at all} and \textit{Almost anything :-)}).

\textbf{Findings}
\begin{itemize}
    \item The most commonly reported refactoring challenge is the starting quality of software. This challenge is closely followed by the difficulty in understanding that software.
    \item Of the tools respondents use today, the most common strengths reported are in automating changes while the most common weaknesses are in usability.
    \item Despite identifying many challenges and weaknesses in today's refactoring tools, 80\% of respondents report having achieved their large-scale refactoring goals.
\end{itemize}

\section{Discussion}
\label{sec:discussion}
Industry software goes through periodic structural changes as part of continuous evolution \cite{iversFSE2020}. While intuitively we know that refactorings support these changes, we also know that they are significantly larger in scale than the kinds of changes common in floss refactoring and so may have different implications on desirable tool support. We discuss the implications of our survey analysis and findings around the need to recognize and study large-scale refactoring, the gaps in tool support for industry's refactoring needs, and the role that tools can play in improving the state of practice. 

\textbf{Recognizing LSR}.
Our findings confirm that large-scale refactoring is a major undertaking that industry software can go through multiple times in its lifetime. Likewise, it confirms that the business consequences of forgoing needed refactoring are often substantial, impacting the ability to deliver features, team productivity and morale, and product quality as seen by users. While existing low-level refactoring knowledge and tool support inform large-scale refactoring, our survey helps identify two distinct characteristics of large-scale refactoring: it is tightly coupled to business needs and the scope of work is broader than local code improvements. 

The following response summarizes these differences clearly:

\begin{itemize}
\item \textit{Agile development practices encourage continuous micro refactorings to make the code a little bit better all the time. ... Refactoring is just part of the job ... like a surgeon washing her hands. ...teams should be continuously refactoring in the small without the need for explicit investment or direction from the wider business. I think larger scale "refactoring" is different in that there is an opportunity cost to doing or not doing the work. It becomes a business decision as to where to invest.}  
\end{itemize}

Large-scale refactoring is a distinct activity for which significant resources need to be allocated, rather than being something that developers can easily weave into their day-to-day work. This context is often not apparent from development artifacts like commit messages or logs of tool use, hampering researchers' ability to study it the way that floss refactoring has been studied. It includes activities like persuading stakeholders of benefits and managing expectations; reasoning activities that span understanding code, requirements, and intent; and integrating data from and capabilities of different tools. In short, it is reasonable to think of it as having a project-like scope that includes building blocks like those of floss refactoring, but also includes many others. As many of these activities are orchestrated across a team (recall that the mean effort estimated for large-scale refactoring from \autoref{fig:lsr_time} is more than 1500 staff days), to study large-scale refactoring and understand the breadth of tool support that is needed, we need to collect and integrate data from more sources.

Large-scale refactoring has architectural implications as demonstrated by the following general feedback responses: 

\begin{itemize}
\item \textit{I believe in such refactorings to be a highly creative task, which often requires with coming up with brand new architectural ideas, or even suggesting actual product changes to make the architecture cleaner ...
I think that large-scale refactorings are more about such things than the code itself.}
\end{itemize} 

\textbf{Tool Support for LSR}.
The finding that developers make little use of existing refactoring tools for large-scale refactoring was unsurprising, as it mirrors studies of smaller-scale refactoring. 
As part of our study, we also sought to understand whether the kinds of tools used in large-scale refactoring differ from those used in other refactoring, the different activities involved in refactoring, and how the tools being used support those activities. Our findings demonstrate that developers typically use multiple tools as part of large-scale refactoring to address the range of activities that are involved. The broad range of tools offered by the respondents go beyond IDEs, including diverse tools such as static code analyzers, issue trackers and wikis, testing tools, and custom scripts. References to custom scripts were significantly more common for large-scale refactoring, suggesting significant tool gaps that are more apparent at scale. Additional research is needed to explore the range of capabilities of such scripts and whether a more general capability can be provided by tool vendors or result from future research.

Responses included several doubts about the feasibility of tool support for large-scale refactoring. A key challenge in developing tools that support large-scale refactoring is improving our understanding of what motivates such activities.  Tools for smaller scale refactoring often start with an assumption that the goal is to remove specific code smells or improve specific code quality metrics. While these improvements offer business value in form of improved software maintainability and developer productivity, these improvements are not always what motivates businesses to invest and hence may be misaligned with project needs. Tools that address different motivations or allow users to express their improvement goals could provide more options for developers.

\textbf{Deciding to Forgo LSR}.
Existing research focusing on floss refactoring has identified barriers and risks that influence decisions about whether to refactor that include: missing resources, risk of introducing an error, difficulty of performing the refactoring, unclear value, constraints set by management, and lack of appropriate tools~\cite{Tempero2017barriers}. 
Our results show similar factors influencing decisions about whether to perform large-scale refactoring. Our respondents reported that prioritizing new features over refactoring and perceiving the cost of refactoring as too high were both the most common and most important reasons that their organizations decided to forgo refactoring. This is not a surprising result to developers in industry or to researchers. 

However, while prioritizing new features over refactoring was the most common reason for forgoing large-scale refactoring, 56\% of respondents reported the inability or slowing pace of delivering features as a consequence of forgoing refactoring. One of our respondents expressed this passionately when asked about the consequences of not refactoring:
\textit{Codebase became shit!} 

Better tools can help change these business decisions and avoid the consequences that follow. As most of the reasons provided boil down to cost-benefit decisions, tools that reduce cost can shift the balance. Our respondents estimated a mean of more than 1500 staff days of effort were spent on large-scale refactorings, suggesting many opportunities to reduce the work involved in refactoring activities. 
While the common wisdom is to focus research on floss refactoring because it is more common (orders of magnitude more so), this perspective neglects the cost difference (large-scale refactorings being orders of magnitude larger).
Researchers and tool vendors would benefit from further research to obtain more granular data on the activities that developers spend the most time on and analyses of which activities they can reliably support with trusted tools.

\textbf{Threats to Validity}.
\label{sec:methodology:threats}
Our threats to validity include: 

\textit{Internal Validity.} Our analysis of survey responses represents a potential threat to internal validity.
To mitigate this threat and to ensure the reliability of our qualitative findings,
we implemented and consistently adhered to established guidelines and best practices for conducting qualitative research,
including comprehensive data use, constant comparison, the use of tables, and refinement of codes through adjudication and
investigator triangulation.

\textit{External Validity.} Our findings are based on the data we collected from 107 survey respondents. We do not make a generalizability claim, but position our findings as observations supported by our data and research literature.  

\textit{Conclusion Validity.} We distributed our survey to a broad audience to collect the most relevant data for our goals. To ensure that we asked the right questions and to avoid introducing our own biases into the wording and selection of questions,
we conducted a series of iterative pilots to identify and address shortcomings in the survey design. Furthermore, we included
several open-ended questions to allow participants to share their views and experiences.

\section{Conclusion}
\label{sec:conclusion}
In order to understand the prevalence, challenges, and tool support for large-scale refactoring we conducted a survey with industry developers. Our analysis of data from 107 respondents, 85\% of whom report to have at least 10 years of experience, confirm that large-scale refactoring is not an unusual occurrence. While floss refactoring is certainly orders of magnitude more common, industry systems undergo multiple large-scale refactorings over their lifetimes and the magnitude of effort involved in each is considerable. Refactoring tools designed to support smaller scale refactoring efforts aren't enough to address the breadth of activities that developers consider a part of large-scale refactoring, and developers encounter a wide range of challenges despite using many different kinds of tools. The anticipated cost of such refactoring, along with business priorities that favor new features, commonly result in organizations forgoing large-scale refactoring, which in turn commonly results in troubling consequences. Our study demonstrates a clear need for better tools and an opportunity for refactoring researchers to make a difference in industry. The results we summarize in this paper is one concrete step towards this goal.

\begin{acks}
This material is based upon work funded and supported by the Department of Defense under Contract No. FA8702-15-D-0002 with Carnegie Mellon University for the operation of the Software Engineering Institute, a federally funded research and development center. References herein to any specific commercial product, process, or service by trade name, trade mark, manufacturer, or otherwise, does not necessarily constitute or imply its endorsement, recommendation, or favoring by Carnegie Mellon University or its Software Engineering Institute. DM21-0915
\end{acks}

\bibliographystyle{ACM-Reference-Format}
\bibliography{knotsurvey}

\end{document}